\documentclass[twocolumn,prb,amsmath,amssymb,showpacs,superscriptaddress,nofootinbib]{revtex4}
\usepackage{graphicx}
\usepackage{dcolumn}
\usepackage{bm}
\usepackage{latexsym}
\usepackage{amstext}
\usepackage{amsxtra}
\usepackage{color}

\definecolor{darkblue}{rgb}{0, 0, 0.8}
\usepackage[colorlinks=true, breaklinks=true, linkcolor=darkblue, citecolor=darkblue, urlcolor=darkblue]{hyperref}

\newcommand{\edr}{aether drag}
\newcommand{\bs}{\boldsymbol}
\usepackage{times}

\begin{document}
\title{Fizeau's ``aether-drag'' experiment in the undergraduate laboratory}
\author{Thierry Lahaye}
\affiliation{Universit\'e de Toulouse, UPS, Laboratoire Collisions Agr\'egats R\'eactivit\'e, IRSAMC; F-31062 Toulouse, France}
\affiliation{CNRS, UMR 5589, F-31062 Toulouse, France}
\author{Pierre Labastie}
\affiliation{Universit\'e de Toulouse, UPS, Laboratoire Collisions Agr\'egats R\'eactivit\'e, IRSAMC; F-31062 Toulouse, France}
\affiliation{CNRS, UMR 5589, F-31062 Toulouse, France}
\author{Renaud Mathevet}
\affiliation{Universit\'e de Toulouse, UPS, Laboratoire Collisions Agr\'egats R\'eactivit\'e, IRSAMC; F-31062 Toulouse, France}
\affiliation{CNRS, UMR 5589, F-31062 Toulouse, France}
\affiliation{Laboratoire National des Champs Magnétiques Intenses, UPR3228 CNRS/INSA/UJF/UPS, Toulouse, France}
\date{\today}

\begin{abstract}
We describe a simple realization of Fizeau's ``aether-drag'' experiment. Using an inexpensive setup, we measure the phase shift induced by moving water in a laser interferometer and find good agreement with the relativistic prediction or, in the terms of $19^{\rm th}$ century physics, with Fresnel's partial-drag theory. This appealing experiment, particularly suited for an undergraduate laboratory project, not only allows a quantitative measurement of a relativistic effect on a macroscopic system, but also constitutes a practical application of important concepts of optics, data acquisition and processing, and fluid mechanics.
\end{abstract}


\maketitle

\section{Introduction}

In introductory courses and textbooks dealing with special relativity, Fizeau's ``aether-drag'' experiment often appears simply as an application of the law of composition of velocities, sometimes in the form of an exercise.\cite{TaylorWheeler1992} However, Albert Einstein himself declared that Fizeau's measurement of the speed of light in moving water was, together with stellar aberration, one of the experimental results that had influenced him most in the development of relativity.\cite{shankland1963} In introductory expositions of Fizeau's experiment, a discussion of the historical development of ideas that lead to it, as well as details about the experimental setup itself, are often lacking. Moreover, many textbooks actually show incorrect experimental arrangements that would not allow in practice for the observation of the effect. Here we show that one can actually perform Fizeau's experiment with rather modest equipment, and that such a project illustrates in an appealing way not only relativistic kinematics, but also interesting aspects of wave optics, data acquisition and processing, and even fluid mechanics.

This article is organized as follows. We first review briefly the historical background of Fizeau's experiment, a ``test'' of special relativity carried out more than half a century before relativity was born! Then, for completeness, we recall in Section~\ref{sec:theory} the derivation of the expected fringe shift in both the relativistic and non-relativistic frameworks, following the usual textbook treatment of the problem. We then turn to the main point of the paper, namely how to reproduce the experiment in an undergraduate laboratory. Section~\ref{sec:setup} is devoted to the description of our apparatus, starting with an emphasis on the experimental trade-offs one needs to address in the design phase. Finally, we discuss in Sec.~\ref{sec:experiment} the results obtained, first with water as a moving medium, and then with air, in order to discriminate between relativistic and non-relativistic predictions. The use of a white-light source instead of a laser is presented in appendix~\ref{app:white:light}, with a discussion of the possible advantages and drawbacks. Appendix~\ref{app:hydraulics} establishes a useful fluid mechanics formula using dimensional analysis.

\section{Historical background}

Since Fizeau's {\edr} experiment is a landmark among the various experimental and theoretical developments leading to special relativity, it is worthwhile to recall briefly the history of these developments. An extensive historical study of the subject is beyond the scope of this paper: in what follows we merely recall the main steps that led to Fizeau's {\edr} experiment, as well as the major subsequent developments.\cite{darrigol}

We begin our reminder in the $17^{\rm th}$ century, at a time when the nature of light was a matter of harsh debate, as evidenced by the famous controversy between Christiaan Huygens and Isaac Newton. The measurement of the speed of light in a material medium of refractive index $n$ was considered a crucial test since Huygens' wave theory implies that the speed of light in the medium is $c/n$, while Newton's corpuscular theory  predicts it to be $nc$, where $c$ is the speed of light in a vacuum, known to be finite since the work of Ole R\"omer in 1676.\cite{romer} Newton's views prevailed until the beginning of the $19^{\rm th}$ century, when interference experiments by Thomas Young and polarization experiments by \'Etienne Malus firmly established the wave theory.

An important step was the measurement by Fran\c{c}ois\ Arago of the deviation of light from a distant star by a prism in 1810.\cite{arago} The idea of Arago is that if the speed of the light coming from distant stars is decreased or increased by the Earth velocity, Newton's theory predicts that the deviation by a prism is different from what would be observed if the source were terrestrial. He therefore tried to detect this difference, with a negative result. It seems to be the first experiment in a long series, which showed the impossibility of detecting the relative motion of light with respect to the Earth.\cite{ferraro2005}

Arago soon became friend with Augustin Fresnel, who had a mathematically sound theory of light waves, and asked him if the wave theory could explain the null result he had found. Fresnel's answer came a few years later.\cite{fresnel} His demonstration is based on the hypothesis of an absolute aether as a support of light waves, associated to a partial drag by transparent media. That is, if the medium of index $n$ moves with speed ${\bs v}$, the aether inside the medium moves only at speed $(1-n^{-2}){\bs v}$. The value of Fresnel's drag coefficient $1-n^{-2}$ precisely gives a null result for the Arago experiment. His demonstration, using some supposed elastic properties of the aether, is however not so convincing by modern standards.\cite{ferraro2005}

The first Earth-based direct measurement of the speed of light was realized in 1849, yet by Hippolyte Fizeau, by means of a rotating cogwheel. This kind of time-of-flight technique was soon improved by L\'eon Foucault who, using a rotating mirror, succeeded in showing that the speed of light is lower in water than in air.\cite{foucault1853} Nevertheless, such \emph{absolute} measurements were far from accurate enough to measure the small change of the speed of light in moving media.

\begin{figure}[t]
\includegraphics[width=80mm]{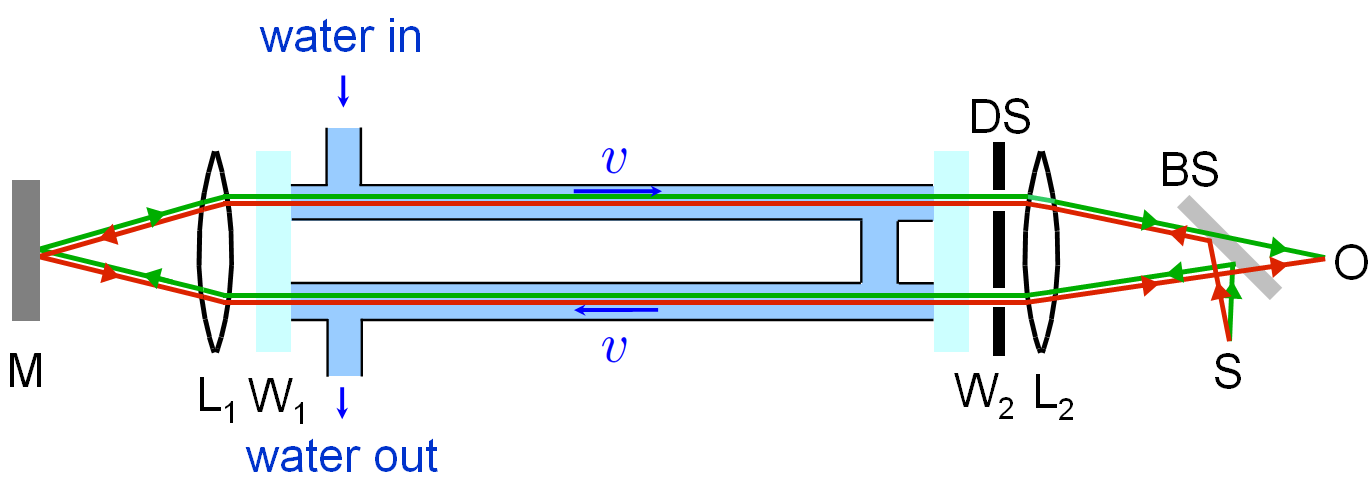}
\caption{(Color online) Sketch of the interferometer used by Fizeau. Figure adapted from.\cite{bookmichelson} For the sake of clarity, the two counter-propagating beams are drawn in different colors. $S$: source; $O$: observer; $M$: mirror; ${W}_i$: windows;  ${L}_i$: lenses; $BS$: beam splitter; $DS$: double slit.}
\label{fig:setup:old}
\end{figure}

This is where Fizeau's aether-drag experiment enters the scene. As we shall see, it is based on a much more sensitive \emph{differential} measurement using the interferometric arrangement shown in Fig.~\ref{fig:setup:old}. It was performed in 1851 and almost immediately reported to the French academy of science, then translated in English.\cite{fizeau1851} He measured an effect in agreement with Fresnel's theory to within a few percent. This unambiguously ruled out concurrent theories postulating total {\edr}.

Many experiments of increasing precision were then undertaken to try to evidence the influence of Earth motion on light propagation, but all gave negative results. It soon became apparent that Fresnel's partial drag prevented to measure any absolute motion of Earth to first order in $v/c$: in what would now be called a review paper,\cite{mascart74} \'Eleuth\`ere Mascart concludes in 1874 (our translation): ``the general conclusion of this memoir would be [...] that the translation motion of the Earth is of no appreciable consequence on optical phenomena produced with terrestrial sources or solar light, that those phenomena do not allow to appreciate the \emph{absolute} motion of a body and that only \emph{relative} motions can be attained.''

Then, in 1881 Albert Michelson designed a new interferometer, which, according to existing theories, could evidence Earth displacement respective to aether because the expected effect was proportional to $(v/c)^2$.  His first measurement was at most half of the expected fringe shift. He then improved the apparatus with Edward Morley. The two physicists gradually became convinced of a negative result. In 1886 they decided to redo Fizeau's experiment, which was the only one with a positive result, and had not been reproduced. With a careful design of the hydraulics part, and an improved design for the interferometer,\cite{michelson1886} they were lead to confirm Fizeau's result, and Fresnel's aether drag, with a much higher precision. However, in their celebrated experiment of 1887\cite{michelson1887} the measured shift was at most 0.01 fringe instead of an expected 0.4. The two experiments were thus incompatible according to existing theories, Fizeau's one needing a partial drag and Michelson-Morley's one a total drag of aether.

History then accelerated. In the late 1880's, George Fitzgerald proposed the length contraction. In 1895, Hendrick Lorentz published his theory of electromagnetic media, in which he derived Fresnel's formula from first principles. At the beginning of the $20^{\rm th}$ century, it became evident that time dilation was also necessary to account for all electromagnetic phenomena. After Albert Einstein published the theory of special relativity in 1905, Max Laue, in 1907, derived Fresnel drag coefficient from the relativistic addition of velocities.\cite{vonlaue1907} All experiments, being either of first (Fizeau) or second (Michelson-Morley) order in $v/c$, were then explained by a single theory with no need for an aether with such special properties.

In the relativistic framework, one can also account for the effects of dispersion, already predicted by Lorentz in 1895. Pieter Zeeman, in his 1914--27 experiments, tried to measure this effect.\cite{zeeman} The supplementary term makes only a few percent correction, but Zeeman succeeded in measuring it.\cite{lerche1977} More recently, the experiment was done in liquids, solids and gases using ring lasers\cite{macek1964} and confirmed the value of the dispersion term to within 15\%.\cite{bilger1972} Fizeau's experiment has also been successfully transposed to neutron matter waves.\cite{klein1981}

As we have seen, Fizeau's {\edr} experiment was a crucial turning point between old and modern conceptions of light and space-time. This therefore makes its replication particulary valuable from a pedagogical point of view.

\section{Theoretical background}
\label{sec:theory}

\begin{figure}[t]
\begin{center}
\includegraphics[width=80mm]{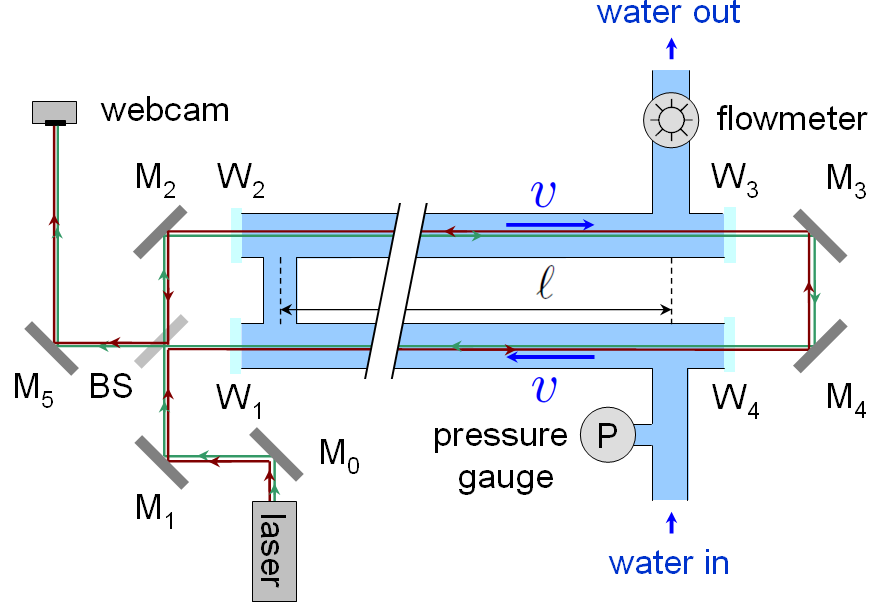}
\end{center}
\caption{(Color online) Sketch of the experimental setup (see text for details). ${M}_i$: mirror; ${W}_i$: windows; $BS$: beam splitter. The two counter-propagating beams are drawn in different colors for clarity.}
\label{fig:setup}
\end{figure}

In this section we recall the derivation of the phase difference $\Delta \varphi$ induced by the motion, at velocity $v$, of the medium of refractive index $n$ in the interferometric arrangement shown in Fig.~\ref{fig:setup}, which is essentially the one used by Michelson and Morley in 1886.\cite{michelson1886} Let us consider first the case where water and monochromatic light, with vacuum wavelength $\lambda$, propagate in the same direction (shown in green on the figure). In the frame where water is at rest, the phase velocity of light is $c/n$. In the laboratory frame, using the relativistic composition of velocities, the phase velocity of light is
\begin{equation}
v_+=\frac{c/n+v}{1+(v\cdot c/n)/c^2}=\frac{c/n+v}{1+v/(nc)}.
\end{equation}
The phase accumulated by light over the propagation distance of $2\ell$ is thus
\begin{equation}
\varphi_+=\frac{2\pi c}{\lambda }\, \frac{2\ell}{v_+}.
\end{equation}
Where light and water propagate in opposite directions (red path on the figure), the corresponding phase $\varphi_-$ is obtained by replacing $v$ by $-v$ in the above result. The phase difference between the two arms of the interferometer thus reads:
\begin{eqnarray}
\Delta\varphi&=&\varphi_{-}-\varphi_+\\
&=&2\pi\frac{2\ell c}{\lambda}\left(\frac{1-v/(nc)}{c/n-v}-\frac{1+v/(nc)}{c/n+v}\right).
\end{eqnarray}
Expanding the above result to first order in $v/c$, we find:
\begin{equation}
\Delta\varphi_{\rm rel.}=2\pi\,\frac{v}{c}\,\frac{4\ell}{\lambda}\left(n^2-1\right).
\label{eq:th:relat}
\end{equation}

It is not difficult to do the same calculation using the non-relativistic addition of velocities $v_\pm=c/n\pm v$. One then finds:
\begin{equation}
\Delta\varphi_{\rm non-rel.}=2\pi\,\frac{v}{c}\,\frac{4\ell}{\lambda} n^2,
\label{eq:th:newton}
\end{equation}
\emph{i.e.} the same functional form except for a coefficient $n^2$ instead of $n^2-1$. In Fresnel's language, this would correspond to a complete aehter-drag. The ratio of the predictions (\ref{eq:th:newton}) to (\ref{eq:th:relat}) is about 2.3 for water ($n=1.33$), and becomes very large ($\sim 1700$) for air ($n-1=3\times 10^{-4}$), whence the interest of performing the experiment also with air (see section \ref{sec:air} below).

As said before, the above derivation is first due to Laue in 1907,\cite{vonlaue1907} and is the one found in most textbooks. It has been pointed out\cite{lerche1977} that such an approach is not rigorous since the relativistic composition of velocities applies to point-like particles, and not to the \emph{phase velocity} of waves. However a rigorous derivation, based on the Lorentz transformation of the four-vector $k^\mu=(\omega/c,{\bs k})$ associated to light, gives the same result provided the light and the medium propagate along the same axis.\cite{lerche1977}

Up to now, we have neglected dispersion, \emph{i.e.} the variation of the refractive index of the moving medium with the light frequency $\omega$. However the frequency of the light in a moving frame is shifted by the Doppler effect. The shifts are opposite for the counterpropagating beams in the interferometer, depicted in red and green in Figs.~\ref{fig:setup:old} and \ref{fig:setup}. They are then subjected to slightly different refraction indices due to dispersion. Then, in Eq.~(\ref{eq:th:relat}) the factor $n^2-1$ has to be replaced by:\cite{jackson}
\begin{equation}
n^2-1+n\omega\frac{{\rm d}n}{{\rm d}\omega}.
\end{equation}
Using the wavelength-dependent refractive index of water found in tables,\cite{water} a simple calculation shows that for water at $\lambda=532~{\rm nm}$, the fringe shift is actually $3.8\%$ greater than what Eq.~(\ref{eq:th:relat}) predicts.

\section{Experimental setup: Fizeau's experiment made easy}
\label{sec:setup}

\subsection{Requirements}

Fizeau's experiment was a real tour de force made possible by the very clever design of the experiment (Fig.~\ref{fig:setup:old}). The improvement by Michelson and Morley essentially transforms the original wavefront-division setup into a much brighter amplitude-division one. In both arrangements, that we would call now Sagnac interferometers,\cite{footnotesagnac} the two interfering beams follow almost exactly the same path (see Fig.~\ref{fig:setup}). This not only doubles the interaction length with the moving medium, but, more importantly, rejects common-mode phase fluctuations (due \emph{e.g.} to turbulence). This arrangement also ensures that the optical path length difference between the two interfering arms is zero when the interferometer is perfectly aligned (see section~\ref{sec:discussion} below).

Equation~(\ref{eq:th:relat}) shows that the expected fringe shift is enhanced by using a short wavelength $\lambda$, and a large product $\ell v$. Let us get an estimate of the requirements on the velocity. We first set $\ell\sim2$~m to make the size of the apparatus reasonable. Second we choose $\lambda=532$~nm that corresponds to cheap diode-pumped solid state lasers. Then one sees that achieving a phase shift on the order of 1~rad with water ($n\simeq1.33$) requires velocities on the order of 4~m$/$s. The experimental setup is thus required to allow for the detection of a shift of a fraction of a fringe, and to produce a water flow of several meters per second.

\begin{figure}[t]
\begin{center}
\includegraphics[width=60mm]{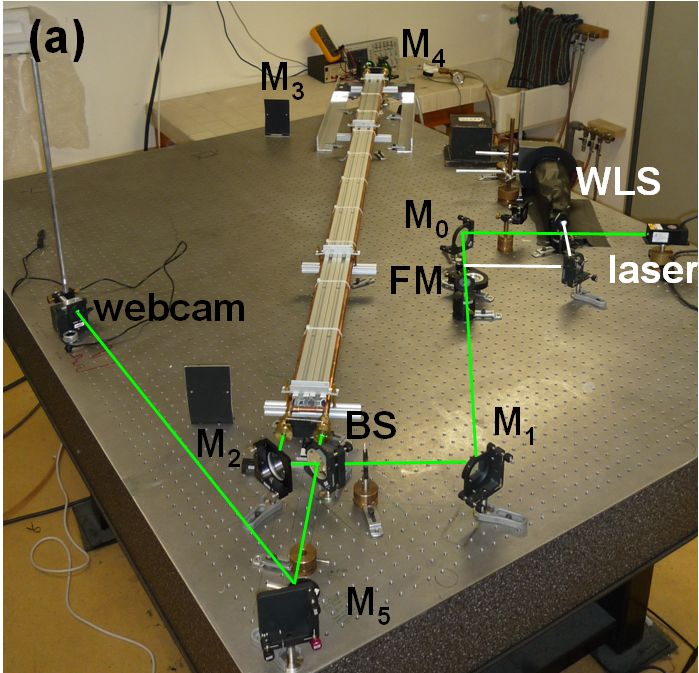}
\includegraphics[width=60mm]{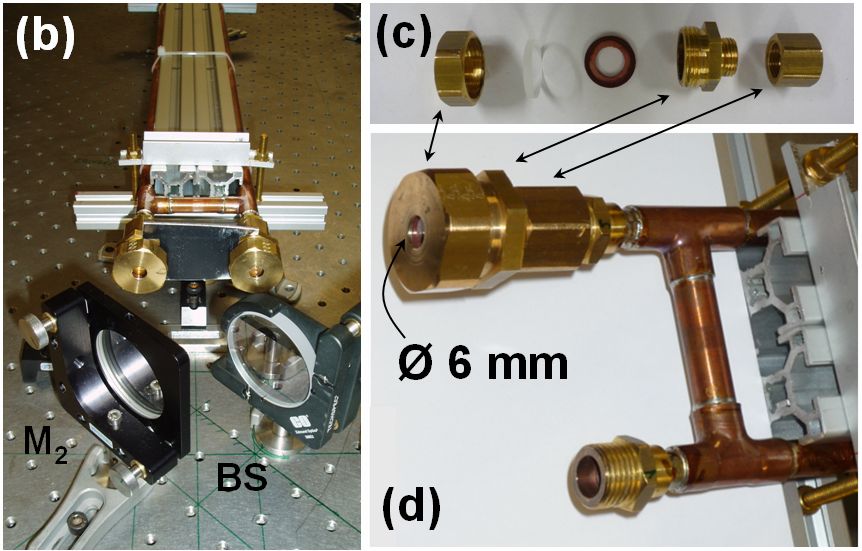}
\includegraphics[width=60mm]{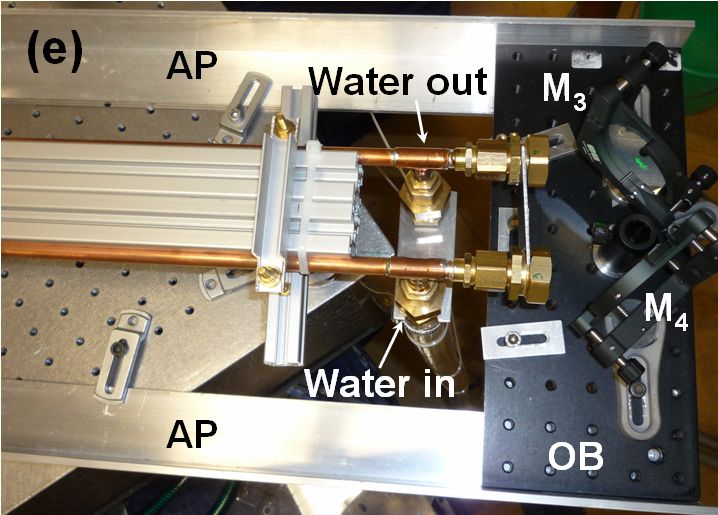}
\end{center}
\caption{(Color online) (a) General view of the experimental setup. WLS: white-light source; FM: flipping mirror. (b) Close-up of one end of the interferometer. (c) The various parts for connecting a window (from left to right: drilled brass BSP blank, BK7 25-mm diameter window, gasket, BSP reducer, BSP adapter). (d) Close-up of the interferometer end shown in (b), with one window disconnected. (e) The other end of the interferometer, showing the water connections and the optical breadboard (OB) supporting mirrors $M_3$ and $M_4$. AP: Aluminum profiles.}
\label{fig:setup:photos}
\end{figure}

The key in the success of the experiment is the care taken in doing the plumbing. We thus describe successively in more details the various components of our experimental setup and refer the reader to the pictures shown in Fig.~\ref{fig:setup:photos}.

\subsection{Hydraulics}

\begin{figure}[t]
\includegraphics[width=80mm]{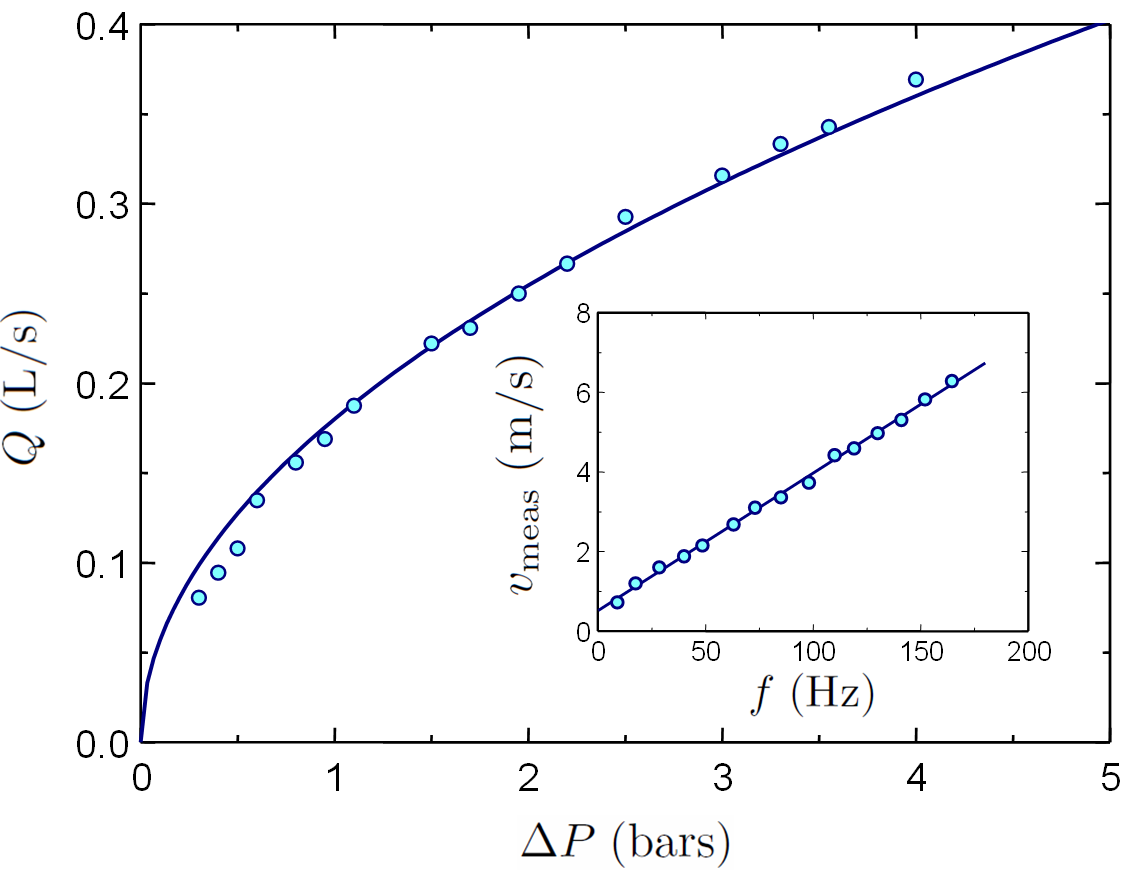}
\caption{(Color online) Flow rate $Q$ through the apparatus as a function of the head loss $\Delta P$. The solid line is a fit by the function $Q=a\sqrt{\Delta P}$ with $a$ as an adjustable parameter; the best fit gives $a\simeq 0.18\;{\rm L/s/\sqrt{bar}}$. Inset: calibration curve of the flowmeter allowing one to infer the water speed $v_{\rm meas}$ from the frequency $f$ of the signal it delivers. The solid line is the result of a linear fit.}
\label{fig:flow}
\end{figure}

For simplicity and low cost, we have built our system from standard piping material available in any hardware store, and in such a way that it can be fed from a regular tap. Ideally a large diameter $d$ of the pipes is desirable. It simplifies the alignment of the interferometer beams and improves the velocity profile flatness over the beam section. However, the volumetric flow rate reads $Q=\pi d^2 v/4$ and increases rapidly with $d$. The typical maximal flow rates available at the water outlets of a laboratory are on the order of ten to twenty liters per minute. To achieve a velocity of $v\sim 5\;{\rm m/s}$, this requires $d \lesssim 10$~mm. One could think, on the contrary, of using small diameters; however, besides being impractical for the alignement of the laser beams, this choice yields increased head loss. The flow rate is then limited by the pressure available from the water distribution system.  The use of a pump to increase the inlet pressure is of little help: in the turbulent regime which is relevant here, the flow rate increases only as the square root of the pressure (see Fig.~\ref{fig:flow} below and Appendix~\ref{app:hydraulics}). In practice, we used 8~mm inner diameter copper tubing, allowing us to reach $v\sim 6\;{\rm m/s}$.

The water pressure $\Delta P$ is varied by opening more or less the tap valve, and measured by a pressure gauge connected to the inlet port (see Fig.~\ref{fig:setup}). We get a continuous measurement of the flow rate $Q$ with a paddlewheel flowmeter.\cite{flowmeter} It delivers a square electric signal whose frequency depends linearly on the flow rate. Its calibration, reported in the inset of Fig.~\ref{fig:flow}, was realized by measuring the volumetric flow rate $Q$ of water through the system using a graduated bucket and a stopwatch. We estimate the accuracy of our crude flow rate calibration to be on the order of 5\%. With moderate effort, a calibration at the percent level or better could certainly be achieved. Note that the velocity $v_{\rm meas}=4Q/(\pi d^2)$ measured in this way is the \emph{mean} velocity, averaged over the radial velocity profile inside the pipes (see Eq.~(\ref{eq:correction}) below) It is not the velocity $v$ appearing in (\ref{eq:th:relat}) which is the one at the position of the beam. We will discuss this point in more details in section~\ref{sec:water}.

\subsection{Mechanics}

The two 2-m long copper pipes are relatively flexible and soft. Their straightness and parallelism is ensured by fastening them on a slotted, $30\times60\;{\rm mm}^2$ cross-section aluminum profile by means of cable ties [see Fig.~\ref{fig:setup:photos}(b)]. The T-shaped connections at their ends [see Fig.~\ref{fig:setup:photos}(d)-(e)] are simply made with low melting point tin solder and a heat gun. On the four ends where the windows need to be installed, male $3/8''$ BSP adapters are soldered.

The windows themselves are 3.3-mm thick, 25~mm-diameter uncoated borosilicate glass substrates.\cite{edmund} They are attached to the pipes via the system shown in Fig.~\ref{fig:setup:photos}(c-d): the window is pressed against a brass female $3/4''$ BSP cap at the center of which a 6-mm diameter hole is drilled. The inner threads were slightly altered with a lathe to fit the window outer diameter. The cap and window are then tightened with a fiber gasket onto a male-male BSP $3/4''$ to $3/8''$ adapter which is itself connected to the male BSP adapter soldered to the pipe via a female-female $3/8''$ BSP adapter.

The one end with hoses connected to the water inlet and outlet in the laboratory sink is extending over the side of the table (see Fig. \ref{fig:setup:photos}(e)). Two L-shaped aluminum profiles screwed on the table support a small piece of optical breadboard on which we mount the two mirrors $M_3$ and $M_4$. In a preliminary set of experiments we tried a configuration in which the water pipes were supported independently from the optical table (and thus from the interferometer) in order to avoid possible detrimental vibrations. However, this was somehow cumbersome, and the much simpler solution of clamping tightly the pipes to the optical table (by means of four regularly spaced post-holders) does not yield any degradation of the measurements.

\subsection{Optics and alignement}

As a light source, we use a cheap diode-pumped, solid state (DPSS) laser delivering a quasi-collimated beam with several milliwatts of light at $\lambda=532$~nm.\cite{laser} The metallic mirrors $M_0$ to $M_4$ and the dielectric beamsplitter are all mounted on kinematic optical mounts. We found it convenient to draw directly the light path on the optical table (see Fig.~\ref{fig:setup:photos}(b)) before precisely mounting the mirrors and the beamsplitter on the optical table. This considerably simplifies the alignment of the interferometer, which is done in the following way.

First, four small diaphragms (diameter $\sim 2$~mm) are positioned just in front of the centers of the windows $W_{1-4}$. The tubing system is then removed, and one walks the beam using the controls of $M_1$, $BS$ and $M_2$ in order to have the beams $W_1W_4$ and $W_2W_3$ passing through the diaphragms. Then, using $M_3$ and/or $M_4$, one aligns the returning beams onto the ingoing ones. In this way, one obtains (possibly after some iterations) a quasi-perfect superposition of the beams, and observes an almost flat intensity profile in the interference field. When tilting slightly one of the mirrors (\emph{e.g.}, $M_3$), nice straight parallel fringes appear.

Now the pipes can be positioned back and clamped onto the table. Water is set to flow, and one makes sure that no air bubble is trapped inside the pipes, especially close to the windows where the diameter is larger. If so, they can be removed by unfastening a little bit the cap while water is flowing.

\begin{figure}[t]
\begin{center}
\includegraphics[width=75mm]{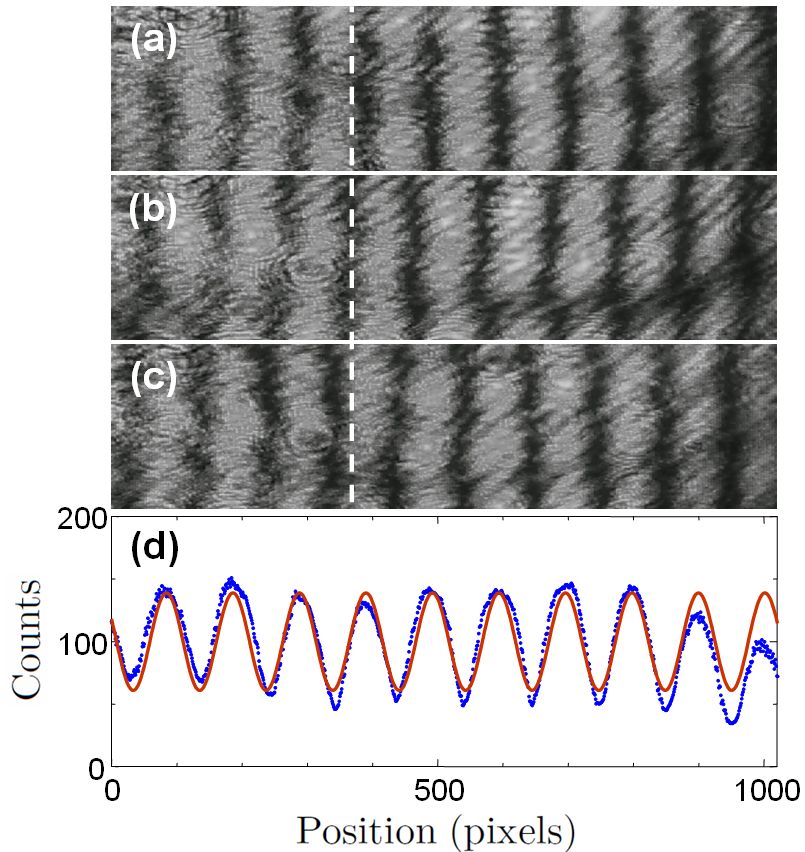}
\end{center}
\caption{(Color online) Sample images of the fringe pattern obtained on the camera for (a) $v=-5.7\;{\rm m/s}$, (b) $v=0.8\;{\rm m/s}$, and (c) $v=5.7\;{\rm m/s}$. The white dashed line shows the position $x_0$ of the central fringe for $v=0$. (d) Processing of the image shown in (c): the dots are the vertically integrated intensity, and the (red) solid line the best fit to~(\ref{eq:fit:func}).}
\label{fig:fringes}
\end{figure}

Instead of the expected fringe pattern, one usually observes caustics and diffuse reflections on the inner sides of the pipes. Indeed, due to the soldering, the parallelism of the windows cannot be ensured. When a beam strikes the window at a small angle from the normal, it is slightly deviated. This deviation is here only partially compensated at the inner glass/water interface. The interferometer has thus to be realigned. After a few iterations, fringes are back (see Fig.~\ref{fig:fringes}).

\section{Experimental results}
\label{sec:experiment}

\subsection{Data acquisition} \label{sec:centralfringe}

For quantitative measurements of the fringe shift, we use an inexpensive webcam\cite{webcam} whose objective lens and IR filter have been removed, in order to expose directly the CMOS detector chip to the fringe pattern. Using the micrometer screws of mirror $M_2$ for instance the fringes are set parallel to one of the axes of the camera chip. From the webcam software the \emph{gamma} correction is set to zero to get a linear response of the detector.\cite{linear} The integration time and light intensity are adjusted to use the full dynamic range of the webcam, taking care not to saturate any pixel.

An important point for later data processing is that, prior to acquiring a series of images, one needs to locate the position of the central fringe on the camera. For this, it is convenient to wobble mirror $M_2$ for example. The fringe spacing varies, and the fringes move symmetrically away from the central one which is dark and does not move. Once the position $x_0$ of the central fringe has been located, the webcam is roughly centered on it, to limit systematic errors due to changes in the fringe spacing (see section \ref{sec:discussion}). The fringe spacing is then adjusted to get about ten fringes on the detector chip. Too few fringes would not allow for an accurate measurement of the fringe position offset and period. On the other hand, if the fringes are too narrow, the resolution of the camera will limit accuracy.

\subsection{Data processing}

We process the images in the following way. We sum up the values of all pixels in a column, and thus obtain a one-dimensional intensity distribution $I(x)$, where $x$ (in pixels) denotes the position along an axis perpendicular to the fringes (Figure~\ref{fig:fringes}). We then fit the data $I(x)$ by the following functional form
\begin{equation}
I(x)=I_0+I_1\sin\left(\frac{2\pi (x-x_0)}{\Lambda}+\Delta\varphi\right).
\label{eq:fit:func}
\end{equation}
Here, $I_0$, $I_1$, $\Lambda$ and $\Delta\varphi$ are adjustable parameters, and $x_0$ is the (fixed) position of the central fringe, determined as explained above.

\subsection{Experimental results with water} \label{sec:water}

\begin{figure}[t]
\includegraphics[width=80mm]{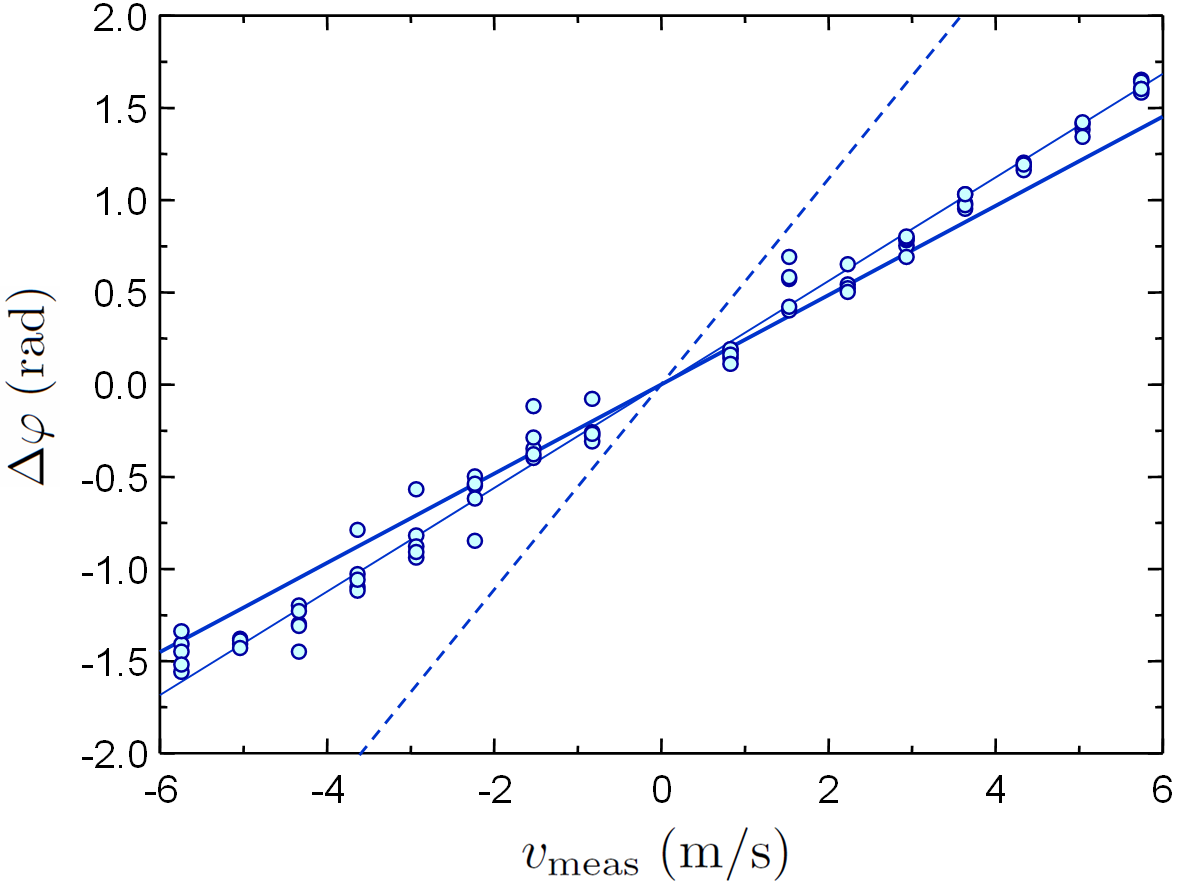}
\caption{(Color online) Experimental results for water as a moving medium. Circles: data; thin solid line: linear fit, giving a slope of $0.274\pm0.003\;{\rm rad\;s/m}$; thick solid line: relativistic theoretical expression~(\ref{eq:th:relat}); dashed line: non-relativistic theoretical expression~(\ref{eq:th:newton}).}
\label{fig:water}
\end{figure}

Acquisition and processing is repeated for various water velocities. Figure~\ref{fig:water} shows the experimentally measured phase difference $\Delta\varphi$ as a function of the water velocity $v$. The origin of phases has been chosen to vanish at zero velocity. As can be seen on the figure, we take five measurements for each velocity in order to increase statistics and get an estimate of
the dispersion of the results. Negative velocities were obtained simply by exchanging the inlet and outlet ports of the tubing system. No points could be recorded for velocities below $1\;{\rm m/s}$. Indeed, when the velocity is low, turbulence in the pipes is not fully developed which leads to low spatial and temporal frequency fluctuations and very unstable pictures. We could not measure the point at zero velocity either, as for this measurement the inlet or outlet valve has to be closed. This produces too few or too much effort on the tubing system which is not stiff enough and light does not get out properly any more. That is why the alignement procedure has to be done, once and for all, but with water flowing in the pipes.

We observe a clear linear dependence of $\Delta\varphi$ on $v_{\rm meas}$. A linear fit (shown as the thin solid line) gives a slope of $0.274\pm0.003\;{\rm rad\;s/m}$. The dashed line is the non-relativistic prediction (\ref{eq:th:newton}), with slope $0.563\;{\rm rad\;s/m}$ which does not match at all the experimental results. The relativistic prediction (\ref{eq:th:relat}), shown as the thick solid line (slope $0.248\;{\rm rad\;s/m}$), is in much better agreement with the experimental data. However, we observe that the experimental points almost systematically lie above the predicted value. This comes from the fact that, as said earlier, we measure the mean velocity $v_{\rm meas}$ averaged over the velocity profile inside the pipes, whereas we need the velocity $v$ appearing in (\ref{eq:th:relat}) which is the one at the position of the beam, \emph{i.e.} on the pipes' axis.

\begin{figure}[t]
\includegraphics[width=80mm]{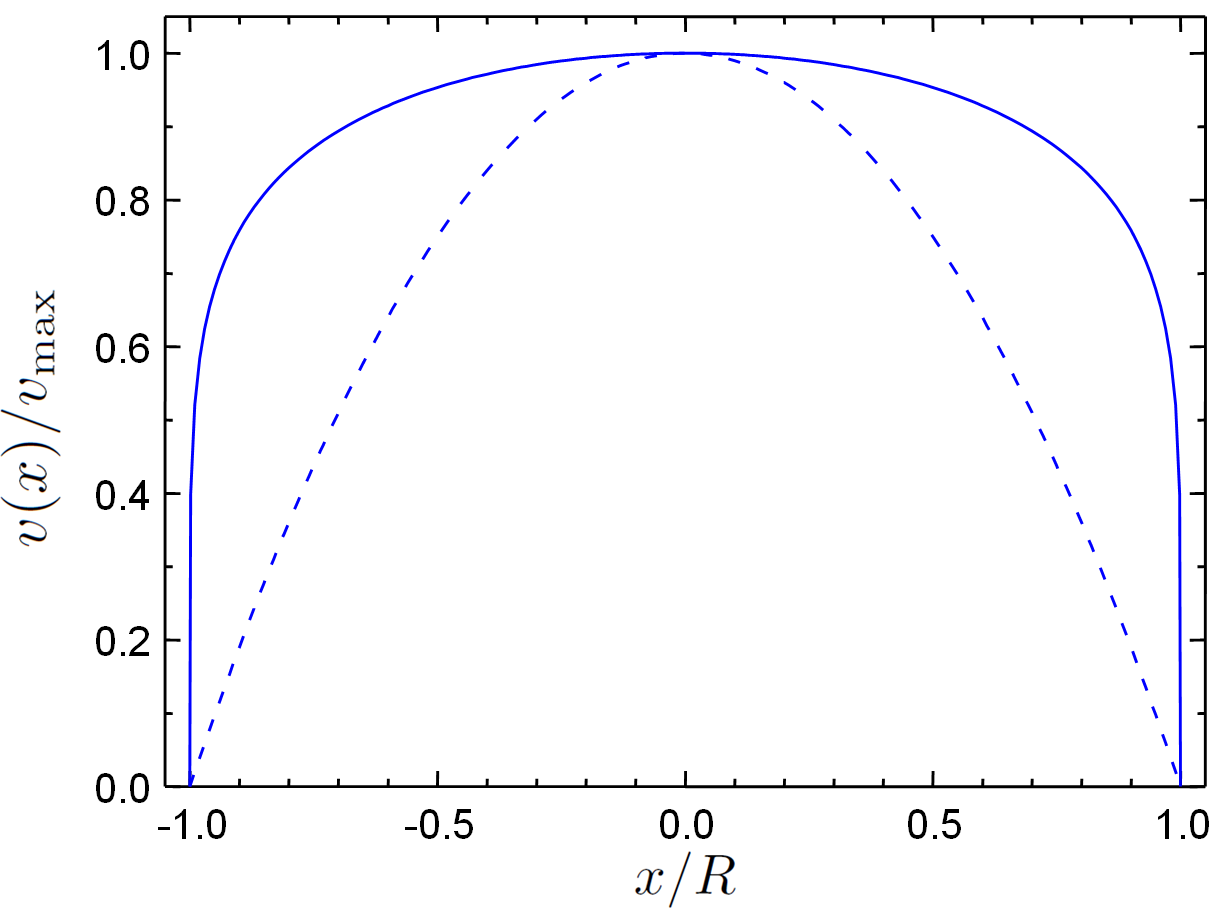}
\caption{(Color online) Dashed line: Poiseuille's velocity profile for laminar flow: $v(r)=v_{\rm max}(1-r^2/R^2)$. Solid line: velocity profile for turbulent flow, modeled here by the empirical form $v(r)=v_{\rm max}(1-r^2/R^2)^{1/6}$.}
\label{fig:profiles}
\end{figure}

Let $v(r)$ denote the radial dependence of the velocity in the pipes of radius $R$ and $v_{\rm max}=v(0)$. As, by construction, the beam is well centered on the pipe one can safely assume $v=v_{\rm max}$. We must therefore multiply the theoretical prediction by the following correction factor
\begin{equation}
\frac{v_{\rm max}}{v_{\rm meas}}=\pi R^2 v_{\rm max}\left/\int_0^R 2\pi r v(r) \,{\rm d}r\right. .
\label{eq:correction}
\end{equation}
A theoretical model for the  radial dependence $v(r)$ is thus required. In the laminar regime (Poiseuille flow), $v(r)$ would have the parabolic shape shown as a dashed line in Fig.~\ref{fig:profiles}, and the correction factor (\ref{eq:correction}) would be equal to 2. But one can check that, for $v\gtrsim 1\;{\rm m/s}$, the Reynolds number ${\rm Re}=vd/\nu$ is already on the order of ${\rm Re}\gtrsim 10^4$ and the flow is turbulent. In the above expression $\nu\sim10^{-6}\;{\rm m^2/s}$ denotes the kinematic viscosity of water. Under these conditions, there is no simple rigorous analytical expression for the velocity profile. However, for the range of Reynolds numbers used here, experimentally measured flow profiles are well reproduced by the empirical law $v(r)=v_{\rm max}(1-r^2/R^2)^{1/6}$,\cite{michelson1886} corresponding to a much flatter velocity profile (see Fig.~\ref{fig:profiles}). Equation~(\ref{eq:correction}) then gives a correction factor of $1.16$. The relativistic prediction (\ref{eq:th:relat}) multiplied by this correction factor, and including also the 3.8\% correction due to dispersion, yields a slope of $0.299\;{\rm rad\;s/m}$ (not shown on Fig.~\ref{fig:water}). The agreement between the experimental and theoretical values is thus at the level of~8\%.

\subsection{Discussion} \label{sec:discussion}

First of all, we conclude that the non-relativistic prediction is clearly ruled out by our measurements. However, the rather good agreement with the relativistic prediction must not be over interpreted. Indeed it is difficult to put a very accurate error bar on the result, as several systematic effects should be studied carefully for such a purpose. First, as stated above, the systematic errors are dominated by our flowrate measurement. A more careful calibration should thus be performed in order to improve the accuracy. Then, the factor of about 1.16 due to the shape of the velocity profile should be measured for our system. A final source of uncertainty is the determination of the actual length $\ell$ appearing in (\ref{eq:th:relat}). In practice, the flow makes a right-angle turn at each end of the pipes. The velocity distribution is affected up and downstream on length scales presumably on the order of the pipe diameter $d$. This implies a correction of order $d/\ell$ (\emph{i.e.}, on the percent level) but, again, an accurate estimation is difficult.

In the end, due to slight distortion of the the tubing when the velocity, and thus the pressure, is varied, the fringe spacing changes a little bit. As said above, an important feature of the Sagnac-like interferometric arrangement used here is that, by construction, it operates at low interference order $p$. This is crucial in order to be sure that when the water is flowing inside the pipes, the observed shift of the fringes does arise from the {\edr} effect and not from a slight change in the fringe spacing. As an example, let us assume that using a different interferometric setup, one observes an interference pattern with 10 fringes, corresponding to interference orders, say $p_1=10^4$ to $p_2=p_1+10$. If, as is very likely, the fringe spacing changes by a quantity as small as $10^{-4}$ in relative value when the water velocity varies, one would observe that our ten fringes would shift, almost as a whole, by as much as one full fringe! We have measured that, for the data presented in the paper, the fringe period $\Lambda$ does not vary by more than 5\% over the full range of velocities, yielding negligible errors due to the low interference orders used here.

\subsection{Experimental results with air} \label{sec:air}

In his original paper,\cite{fizeau1851} Fizeau states that he performed the experiment with air as a moving medium and that (our translation) ``the motion of the air does not produce any sensible displacement of the fringes'', in agreement with the partial drag prediction (\ref{eq:th:relat}). On the contrary, the non-relativistic equation~(\ref{eq:th:newton}) predicts a measurable shift.

It is thus interesting to perform the experiment also with air. We do so by using a standard compressed air outlet, as available in most laboratories. One actually needs very moderate pressures in order to achieve relatively high velocities for the air flow in the $d=8$~mm pipes: only 0.2~bar typically yields $v_{\rm meas}\simeq 35\;{\rm m/s}$. Measuring the air velocity is not as straightforward as with water; we found it convenient to use a hot-wire anemometer\cite{anemometer} placed in a $D=18$~mm inner diameter pipe at the outlet of the $d=8$~mm pipes. The velocity $v_{\rm meas}$ in the interaction region of the interferometer is then deduced from the measured velocity $v_{\rm anem}$ at the anemometer position via volumetric flow conservation $v_{\rm meas}=v_{\rm anem}(D/d)^2$. This assumes incompressible flow, which is valid since the air velocity is much smaller here than the speed of sound.\cite{compressibility}

\begin{figure}[t]
\includegraphics[width=80mm]{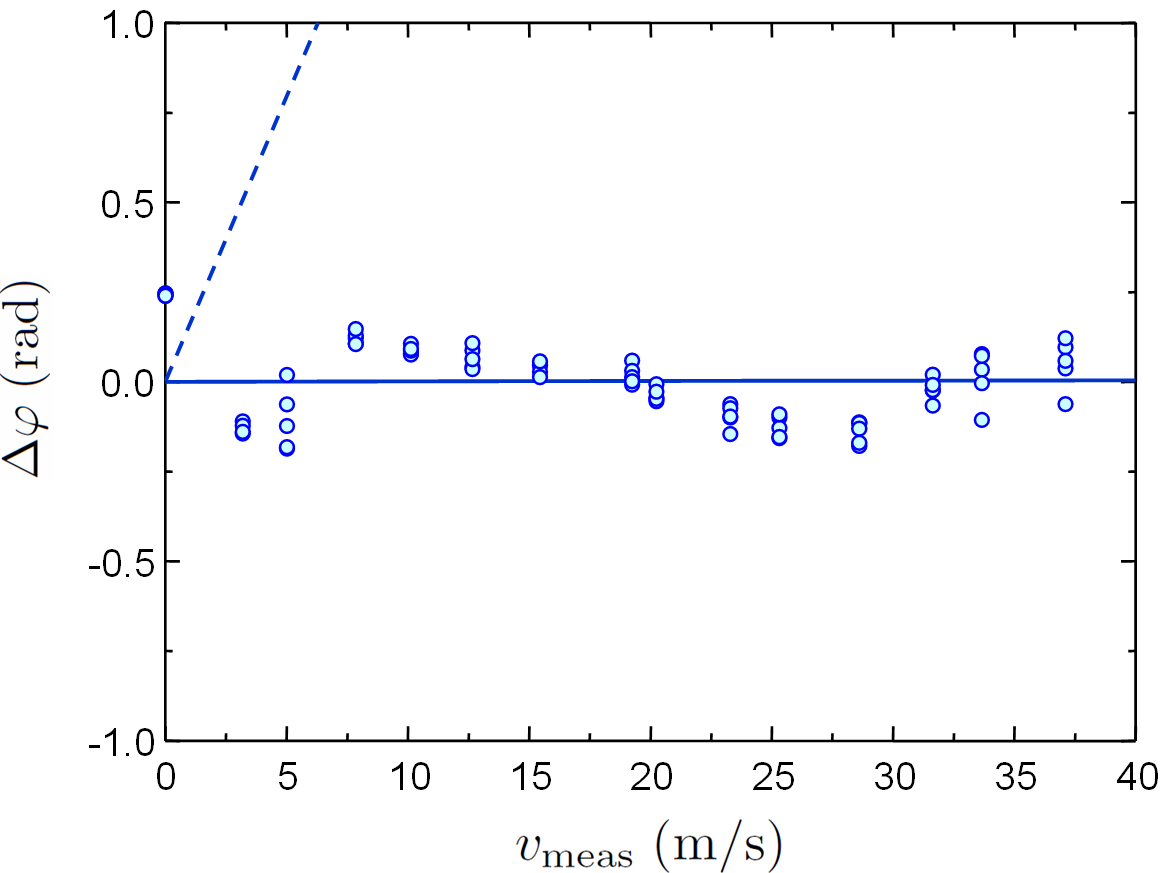}
\caption{(Color online) Experimental results for air as a moving medium. Circles: data; solid line: relativistic prediction~(\ref{eq:th:relat}); dashed line: non-relativistic prediction~(\ref{eq:th:newton}).}
\label{fig:air}
\end{figure}

Figure~\ref{fig:air} shows the measured phase shifts (circles) along with the predictions of equation (\ref{eq:th:relat}) and (\ref{eq:th:newton}). We could not clearly identify the reason(s) behind the seemingly oscillating behavior of the measured fringe shift with velocity. In any case, the non-relativistic prediction is clearly ruled out by the measurements, which are, on the contrary, compatible with the relativistic calculation.

\section{Conclusion and outlook}
\label{sec:conclusion}

Using rather modest equipment, we have shown that Fizeau's ``aether-drag'' experiment can be reproduced in the undergraduate laboratory at a quantitative level. It not only makes a nice practical introduction to the sometimes abstract concepts of special relativity, but also constitutes an interesting application of several branches of experimental physics.

Immediate improvements of the setup described in this paper would consist in (i) calibrating the flowrate more carefully, and (ii) increasing the stiffness of the tubing system. A natural extension of this work, suitable for a long-term student project, would consist in trying to study in details systematic effects, for instance the determination of the effective length $\ell$. A possible way to measure this systematic effect could be to start from the full pipe length and then repeat the experiment for shorter and shorter pipe lengths. The effect can then be evaluated measuring the dependence of the slope $\Delta\varphi/v$ on the pipe length.

A more ambitious extension, suitable for advanced undergraduates, would illustrate more modern optical techniques. For instance, one may use a ring cavity of moderately high finesse (say $\mathcal{F}\sim 100$ to $1000$) and measure the variation of the resonance frequencies of the two counterpropagating modes when the velocity of the medium is varied. A gain in sensitivity by a factor $\mathcal{F}$ is then expected. Such techniques, with ultra-high finesse cavities, are currently used to measure \emph{e.g.} non-reciprocity effects in the propagation of light with amazing sensitivities.\cite{cecile}

\begin{acknowledgements}
We thank \'Eric Desmeules for being at the origin of this work and his students M\'elodie Andrieu and Laurane Boulanger for help in setting up a preliminary version of the apparatus during their `TIPE' project: the data shown in Figs.~\ref{fig:flow} and \ref{fig:water} were essentially acquired by them. We thank Jacques Vigu\'e for useful discussions. David Gu\'ery-Odelin created the conditions that made this project possible.  R. M. dedicates his contribution to Jos\'e-Philippe P\'erez for inspiring discussions over the years. Funding by CNRS is acknowledged.
\end{acknowledgements}

\appendix

\section{Using a white-light source instead of a laser}
\label{app:white:light}

\begin{figure}[t]
\includegraphics[width=70mm]{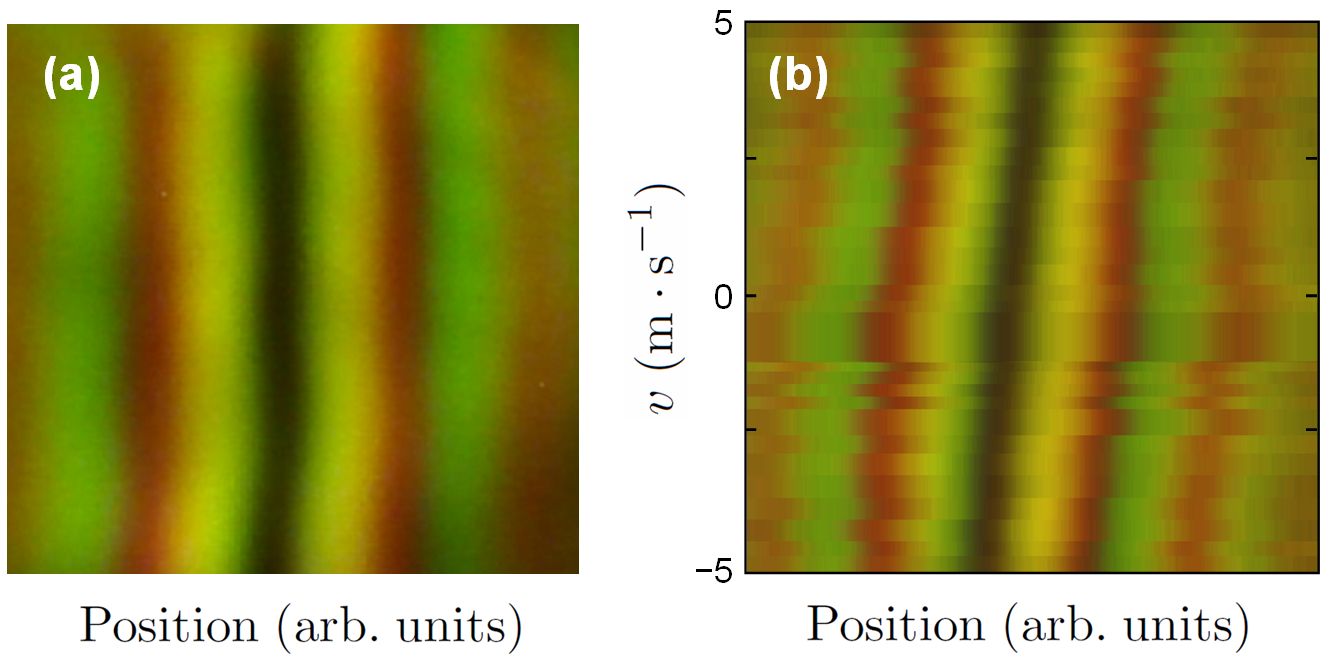}
\caption{(Color online) (a) A sample white-light fringe pattern. (b) Composite image of 22 fringe patterns obtained for different water velocities $v$. One clearly observes the linear shift of the central fringe position as a function of $v$.}
\label{fig:white:light:fringes}
\end{figure}

We have also performed the experiment using a white-light source instead of a laser. The source is a 1-mm diameter iris illuminated by a 55~W halogen lamp (of the type used for car headlights) and a condenser lens. The resulting diverging beam is collimated by a 100-mm focal length lens, and superimposed onto the path of the laser beam using two mirrors. The second one, located between $M_0$ and $M_1$, is a flipping one so that one can switch easily between the laser and the white light source (see Fig.~\ref{fig:setup:photos}(a)). Once the interferometer has been aligned with the laser, white-light fringes are readily observed. Naturally, if the iris is opened the luminosity is increased at the expense of spatial coherence. The contrast of picture is lost. The fringes are then localized in the vicinity of mirror $M_3$. A colorful image as Fig.~\ref{fig:white:light:fringes}(a) is then recovered with a converging lens that conjugates $M_3$ and the detector plane.

The advantages of using a white source is that (i) the position of the dark central fringe can be found without ambiguity as the contrast of the colored fringes vanishes rapidly away from the zero path-length difference, and (ii) as compared to using a laser is that unwanted interference fringes, due to scattering on dust particles for instance, as well as speckle, are suppressed. There are however a certain number of drawbacks. Besides the reduced luminosity, making quantitative comparisons with theory is obviously much more difficult than in the monochromatic case. Indeed, one would need to measure the light spectrum, as well as the wavelength-dependent reflectivity (including phaseshifts) introduced by the beamsplitter in order to model quantitatively the fringe pattern. For instance, we have observed indeed that using a metallic beamsplitter instead of a the dielectric one alters significantly the colors and the contrast of fringe pattern obtained in white light.

We made a composite of 22 images taken for various water velocities $v$. The result, shown on Fig.~\ref{fig:white:light:fringes}(b)), clearly shows that the fringe position shift linearly with velocity. However, as explained above, a quantitative analysis of such an image is not easy, and the motivation behind this figure is more of an esthetic character.

\section{Turbulent head loss in a circular pipe: dimensional analysis approach} \label{app:hydraulics}

In the standard introductory physics curriculum, the computation of the head loss in a circular pipe is done using Poiseuille's equation, valid for laminar flow. It is much less frequent to present te case of turbulent flow to undergraduate students. Reference to the Moody diagram giving the so-called \emph{friction factor} as a function of the Reynolds number (and pipe roughness) can be found in engineering-oriented textbooks, but may appear as quite involved to beginning physics students. In this small appendix, we show how simple dimensional analysis can be used to infer a plausible expression for the turbulent head loss, at least its dependence on flow rate and pipe diameter, two parameters that are crucial for the design of our experimental setup.

We consider the head loss $\Delta P$ for the flow of a fluid of density $\rho$ and kinematic viscosity $\nu$ across a circular pipe of diameter $d$ and length $l$, flowing with a volumetric flow rate $Q$. Let us make two assumptions. First, for an infinitely long pipe, only the pressure gradient $\Delta P/l$ is physically relevant. Moreover, the limit of a very large Reynolds number corresponds formally to the limit $\nu\to 0$. Then, in that case, the viscosity $\nu$ should not appear explicitly in the expression for the head loss. Under these conditions, we expect the following functional form for the head loss to hold:
\begin{equation}
\frac{\Delta P}{l} =A \rho^\alpha Q^\beta D^\gamma,
\end{equation}
where $A$ is a dimensionless constant and $(\alpha,\beta,\gamma)$ exponents to be determined. Equating the dimensions of both sides of the above equation yields three equations for the exponents, giving in the end:
\begin{equation}
\frac{\Delta P}{l} =A \frac{\rho \, Q^2}{D^5}.
\end{equation}
This expression agrees well with empirical formulae used in an engineering context if one chooses $A\sim3\times 10^{-2}$. This is typically what one would find using the friction factor obtained in Moody diagram\cite{guyon,faber} for our typical Reynolds numbers. The value of the coefficient $a$ obtained in fitting the $Q(\Delta P)$ data of Fig.~\ref{fig:flow} yields $A\simeq2.5\times 10^{-2}$, in good agreement with the previous estimate.


\begin{thebibliography}{99}
\bibitem{compressibility}
Corrections due to the compressibility of air are on the order of $(v/v_{\rm s})^2$, where $v$ is the flow velocity and $v_{\rm s}\simeq340\;{\rm m/s}$ the speed of sound in air. For our parameters, it thus amounts to about 1\% at most, and is completely negligible with respect to other sources of uncertainty.
\end{thebibliography}
\end{document}